\documentclass[conference]{IEEEtran} 
\usepackage{graphicx}
\usepackage{subfigure}

\usepackage{bm}
\usepackage{amsfonts}
\usepackage{amsthm}
\usepackage{amssymb}

\usepackage{color}
\usepackage{mathrsfs}

\usepackage[utf8]{inputenc} 

\usepackage[outdir=./]{epstopdf}

\usepackage{cite}

\usepackage{booktabs}

\newtheorem{thm}{Theorem}
\newtheorem{lemma}{Lemma}

\ifCLASSINFOpdf
\else
\fi
\usepackage{amsmath}
\usepackage{amsthm}

\usepackage{algorithm}
\usepackage{algorithmic}
\ifCLASSOPTIONcompsoc
 \usepackage[caption=false,font=normalsize,labelfont=sf,textfont=sf]{subfig}
\else
 \usepackage[caption=false,font=footnotesize]{subfig}
\fi
\usepackage{dblfloatfix}
\begin{document}
%
\title{RIS-assisted Integrated Sensing and Communications: A Subspace Rotation Approach}
%
%
%


\author{
\textit{(Invited Paper)}\\
Xiao Meng$^{1,2}$, Fan Liu$^{2}$, Shihang Lu$^{2}$, Sundeep Prabhakar Chepuri$^{3}$ and Christos Masouros$^{4}$  
\thanks{Corresponding author: Fan Liu (email: liuf6@sustech.edu.cn)}\\ 
$^{1}$ Beijing Institute of Technology, Beijing, China \\
$^{2}$ Southern University of Science and Technology, Shenzhen, China\\
$^{3}$ Indian Institute of Science, Bangalore, India \\
$^{4}$ University College London, London, UK\\
}

\maketitle

\begin{abstract}
In this paper, we propose a novel joint active and passive beamforming approach for integrated sensing and communication (ISAC) transmission with assistance of reconfigurable intelligent surfaces (RISs) to simultaneously detect a target and communicate with a communication user. 
We first show that the sensing and communication (S\&C) performance can be jointly improved due to the capability of the RISs to control the ISAC channel. 
In particular, we show that RISs can favourably enhance both the channel gain and the coupling degree of S\&C channels by modifying the underlying subspaces.
In light of this, we develop a heuristic algorithm that expands and rotates the S\&C subspaces that is able to attain significantly improved ISAC performance.
To verify the effectiveness of the subspace rotation scheme, we further provide a benchmark scheme which maximizes the signal-to-noise ratio (SNR) at the sensing receiver while guaranteeing the SNR at the communication user. 
Finally, numerical simulations are provided to validate the proposed approaches.
\end{abstract}

\begin{IEEEkeywords}
ISAC, RIS, beamforming, subspace.
\end{IEEEkeywords}

%
\IEEEpeerreviewmaketitle

\section{Introduction}
%
%
%
%
\IEEEPARstart{S}{ensing} has been regarded as an important function in the next-generation wireless networks\cite{Saad6G,liu2022seventy}. Many emerging mobile applications, such as smart manufacturing and vehicle to everything, not only require high-quality communication with low latency and high rate, but also require location information with high precision\cite{9606831}. 
To provide better performance and to efficiently use the spectrum, energy, and hardware, integrating the sensing functionality and communication into a single system becomes a promising approach. By sharing the hardware and wireless resources and jointly designing the waveform and signal processing flow between S\&C, a significant performance gain can be obtained in integrated sensing and communication (ISAC) systems\cite{liu2021integrated,9540344}.

In parallel to the ISAC technology, reconfigurable intelligent surfaces (RISs) or intelligent reflecting surfaces (IRSs), which are well known for its ability to modify the wireless propagation environment, has also drawn significant attention from both academia and industry\cite{8811733,9779545,8910627}. By designing the phase shift matrix, RIS is capable of simultaneously modifying the communication channel and the sensing channel, which is favorable for an ISAC system\cite{8910627,liu2021integrated,9909562,9593143,9729741,9416177,9591331,9769997,SankarRIS1}.
In particular, RIS can be designed to diminish interference between the radar and communication system\cite{9729741},
and may also be designed to reduce the multi-user interference (MUI)\cite{9416177,9591331}.
As a step forward, jointly designing the RIS and transmit beamformer, one may leverage the constructive interference to facilitate the ISAC transmission\cite{9769997}.

\begin{figure}[t]
  \centering
  \includegraphics[width=0.30\textwidth]{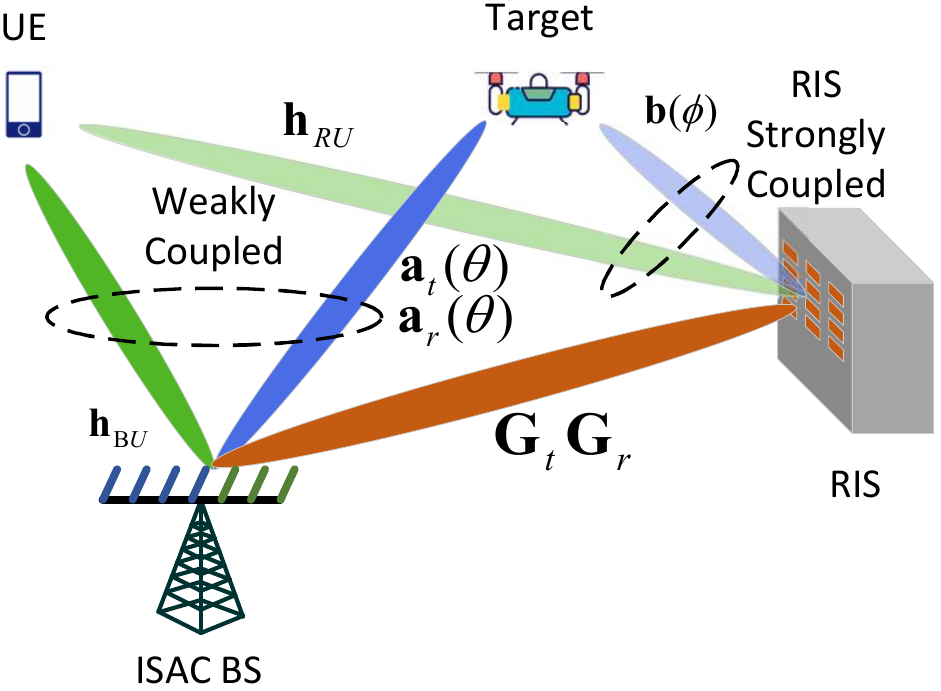}
  \caption{RIS-assisted ISAC system model.}\label{Demo}
  \end{figure}

Motivated by the above research, in this paper we investigate the joint active and passive beamforming design for the RIS-assisted ISAC system, where a multi-antenna base station (BS) simultaneously serves a single antenna user and tracks a target. 
We first point out that compared with the individual S\&C systems, the additional performance gain provided by the RIS mainly comes from the improvement of channel/subspace correlation and the enhanced channel gain, by presenting a brief analysis on the RIS-assisted channel as well as the structure of the beamformer.
Based on these findings, we then develop a heuristic method to rotate and expand the S\&C subspaces.
To provide a performance baseline, we also introduce a benchmark beamforming technique to maximize the sensing signal-to-noise ratio (SNR) while guaranteeing the communication SNR.
To solve the optimization problem, we employ alternative optimization (AO) algorithm to iteratively optimize the active beamformer at the BS and passive beamformer at the RIS.
Finally, we provide numerical results to verify the effectiveness of the proposed subspace rotation approach.

\section{System Model}
Let us consider an RIS-assisted ISAC system, where a multi-antenna BS simultaneously serves a single-antenna user equipment (UE) and tracks a single target.
As shown in Fig. \ref{Demo}, the S\&C channels may be weakly coupled, resulting in poor performance of the ISAC system (which will be detailed later). 
An RIS with $M$ elements is deployed to provide additional strongly-coupled channels, thus improving the joint performance.
The BS is equipped with $N_t$ transmit antennas and $N_r$ receive antennas, which transmits an ISAC signal $\mathbf{x}(t)$ to perform both S\&C tasks.
By denoting the communication channel, the transmit sensing channel, and the receive sensing channel as $\mathbf{h}_c \in \mathbb{C}^{N_t }$, $\mathbf{h}_t \in \mathbb{C}^{N_t }$, and $\mathbf{h}_r \in \mathbb{C}^{N_r }$, respectively, the signal model of this ISAC system can be expressed as 
\begin{align}
  &\text{Sensing Model:}\; \mathbf{y}_s(t) = \mathbf{h}_r^* \mathbf{h}_t^H \mathbf{x}(t)  + \mathbf{z}_s, \forall t,\\ 
  &\text{Comms Model:}\; y_c(t) = \mathbf{h}_c^H \mathbf{x}(t) + z_c, \forall t,
\end{align}
where $\mathbf{x}(t) = \mathbf{w} s(t)$ with $\mathbf{w} \in \mathbb{C}^{N_t }$ denoting the ISAC beamformer and $s(t)$ denoting the communication signal with unit power, $\mathbf{y}_s \in \mathbb{C}^{N_r}$ and $y_c$ respectively denote the received signal at the BS and the UE. Here, $\mathbf{z}_s \in \mathbb{C}^{N_r}$ denotes the additive white Gaussian noise (AWGN) vector at the BS with the variance of each entry being $\sigma_s$ and $z_c$ denotes the AWGN at the UE with the variance being $\sigma_c$.
In particular, the channel can be modeled as
\begin{align}
  &\mathbf{h}_{t}\! =\! \alpha_t \mathbf{a}_t(\theta)\! +\! \alpha_g\mathbf{G}_{t}\bm{\Phi}\mathbf{b}(\phi) \approx   \alpha_t( \mathbf{a}_t(\theta) + \mathbf{G}_{t}\bm{\Phi}\tilde{\mathbf{b}}(\phi)),\label{Tx_Channel}\\
  &\mathbf{h}_{r}\! =\! \alpha_r\mathbf{a}_r(\theta)\! +\! \alpha_g\mathbf{G}_{r}\bm{\Phi}\mathbf{b}(\phi)   \approx  \alpha_r(\mathbf{a}_r(\theta) + \mathbf{G}_{r}\bm{\Phi}\bar{\mathbf{b}}(\phi)),\label{Rx_Channel}\\
  &\mathbf{h}_c =  \mathbf{h}_{BU} + \mathbf{G}_t\bm{\Phi}\mathbf{h}_{RU},\label{Comms_Channel}
\end{align}
where $\alpha_t$, $\alpha_r$, and $\alpha_g$ denote the reflection coefficient and path-loss coefficient from the transmit antenna to the target, that from the target to the receive antenna and that from the RIS to the target, respectively, $\mathbf{a}_t \in \mathbb{C}^{N_t}$, $\mathbf{a}_r \in \mathbb{C}^{N_r}$ and $\mathbf{b} \in \mathbb{C}^{M }$ denote the steering vector from the transmit antenna and receive antenna to the target and the steering vector from the RIS to the target, $\mathbf{G}_t \in \mathbb{C}^{N_t\times M}$ and $\mathbf{G}_r \in \mathbb{C}^{N_r \times M}$ denote the channel from the transmit antenna array and the receive antenna array to the RIS, $\mathbf{h}_{BU} \in \mathbb{C}^{N_t}$ and $\mathbf{h}_{RU} \in \mathbb{C}^{M}$ denote the channel from the BS to the UE and that from the RIS to the UE and $\bm{\Phi} \in \mathbb{C}^{M \times M}$ is the diagonal phase shift matrix of the RIS.
While $\alpha_t$, $\alpha_r$, and $\alpha_g$ may be challenging to be explicitly obtained, their relationship can be approximately estimated by leveraging the geometric relationship among the BS, the RIS, and the target. For notational convenience, we normalize $\mathbf{b}$ to $\tilde{\mathbf{b}}$ and $\bar{\mathbf{b}}$ in (\ref{Tx_Channel}) and (\ref{Rx_Channel}), respectively.

To provide better sensing performance while ensuring communication quality, we maximize the sensing SNR with a given communication SNR threshold by jointly optimizing the phase shift matrix $\bm{\Phi}$ and beamformer $\mathbf{w}$, in which case the optimization problem can be formulated as
\begin{subequations} \label{Opt_Raw}
  \begin{align}
    \max_{\bm{\Phi},\mathbf{w}} \quad &\text{SNR}_s \\
    s.t. \quad\ &\text{SNR}_c \geq \Gamma_0,\\
                &||\mathbf{w}||^2 \leq P_t,\\
                &|\varphi_m| = 1, \forall m = 1,2,...,M,\\
                &\bm{\Phi} = \text{diag}(\varphi_1,....,\varphi_M),
  \end{align} 
\end{subequations}
where  $\text{SNR}_s = \Vert\mathbf{h}_r^*\mathbf{h}_t^H\mathbf{w}\Vert^2/\sigma_s^2$ denotes the sensing SNR, $\text{SNR}_c = |\mathbf{h}_c^H\mathbf{w}|^2/\sigma_c^2$ denotes the communication SNR, $\varphi_m$ denotes the $m$-th non-zero element of the diagonal phase shift matrix $\bm{\Phi}$, $\Gamma_0$ denotes the communication SNR threshold, and $P_t$ denotes the transmit power budget.

\section{ Proposed Subspace Rotation Scheme}
In this section, we show that by introducing RIS we can improve the performance of an ISAC system over that of individual S\&C systems from the channel subspace perspective.
Towards this end, we propose a novel algorithm by exploiting the subspace correlation between S\&C channels.

\begin{figure}[t]
  \centering
  \includegraphics[width=0.30\textwidth]{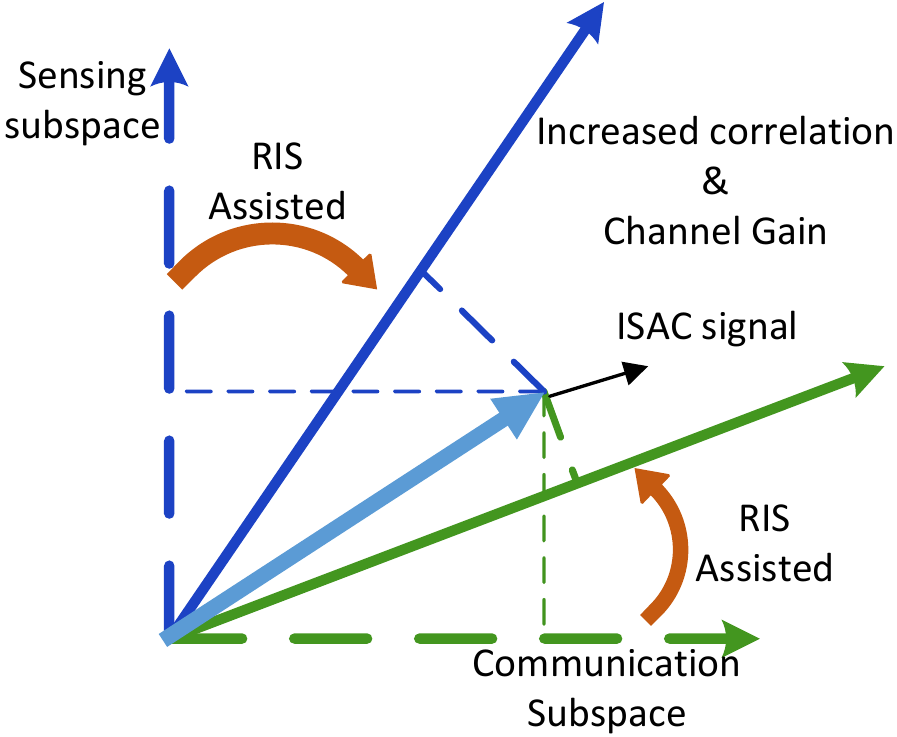}
  {\caption{ISAC performance enhanced by RIS: subspace expansion \& rotation.}}
  \label{Subspace_Demo}
\end{figure}

\subsection{ISAC Beamforming Design Without RIS}
To begin with, we first investigate the optimal beamformer without the assistance of RIS, in which case we have $\mathbf{h}_t = \alpha_t \mathbf{a}_t(\theta)$, $\mathbf{h}_r = \alpha_r \mathbf{a}_r(\theta)$ and $\mathbf{h}c = \mathbf{h}_{BU}$.
To maximize the sensing SNR while guaranteeing the communication quality, the optimization problem can be formulated as
\begin{subequations} \label{Opt_prob_w}
  \begin{align}
    \max_{\mathbf{w}} \quad &\text{SNR}_s \\
    s.t. \quad\ &\text{SNR}_c \geq \Gamma_0,\\
                &\|\mathbf{w}\|^2 \leq P_t.
  \end{align} 
\end{subequations}

\begin{lemma}
  The optimal solution of (\ref{Opt_prob_w}) satisfies 
  \begin{equation}
    \mathbf{w} \in \operatorname{span} \{ \mathbf{h}_c, \mathbf{h}_t \}.
  \end{equation}
  Proof. \textup{See \cite{liu2021cramer}}.
  \end{lemma} 
  This indicates that the optimal solution always belongs to the linear subspace spanned by the communication subspace $\mathbf{h}_c$ and the sensing subspace $\mathbf{h}_t$. Using Lemma 1, the optimal solution for (\ref{Opt_prob_w}) is given in the following theorem.
  \begin{thm}\label{Thm_optimal_beamformer}
    The optimal solution to (\ref{Opt_prob_w}) is 
    \begin{subequations}
    \begin{align}\label{Opt_w}
      \mathbf{w} = 
      \begin{cases}
       \sqrt{P_t} \frac{\mathbf{h}_t}{\|\mathbf{h}_t\|}, \; \textup{if} \; P_t |\mathbf{h}_c^H \mathbf{h}_t|^2 \geq \Gamma_0 \sigma_c^2 \|\mathbf{h}_t\|^2\\
       x_1 \mathbf{u}_1 + x_2 \mathbf{u}_2, \; \textup{otherwise},
      \end{cases}
    \end{align}
    where 
    \begin{equation}
      \mathbf{u}_1 = \frac{\mathbf{h}_c}{\|\mathbf{h}_c\|}, \quad 
      \mathbf{u}_2 = \frac{\mathbf{h}_t - (\mathbf{u}_1^H \mathbf{h}_t) \mathbf{u}_1 }{\|\mathbf{h}_t - (\mathbf{u}_1^H \mathbf{h}_t) \mathbf{u}_1\|},
    \end{equation}
    \begin{equation}
      x_1 = \sqrt{\frac{\Gamma_0 \sigma_c^2}{\|\mathbf{h}_c\|^2}}  \frac{\mathbf{u}_1^H\mathbf{h}_t}{|\mathbf{u}_1^H\mathbf{h}_t |}, \quad
      x_2 = \sqrt{P_t - \frac{\Gamma_0 \sigma_c^2}{\|\mathbf{h}_c\|^2}} \frac{\mathbf{u}_2^H\mathbf{h}_t}{|\mathbf{u}_2^H\mathbf{h}_t |}.
    \end{equation}
  \end{subequations}
    Proof. \textup{ See \cite{liu2021cramer}.}
  \end{thm}



\subsection{Subspace Expansion and Rotation in ISAC Channels}
\textbf{Subpsace Expansion:}
By examining (\ref{Tx_Channel})-(\ref{Comms_Channel}), it is easy to observe that the presence of the RIS provides extra propagation paths. Accordingly, the first positive effect provided by the RIS is the additional channel gain for both S\&C, which is equivalent to expanding the S\&C subspace.

\textbf{Subspace Rotation:}
RIS also offers another promising way to improve the ISAC performance. By adjusting the phase shifters, RIS is capable of artificially rotating the S\&C subspace and thus increasing the channel correlation between them, in which case signal power may be reused appropriately by the dual functionalities, improving the resource efficiency.

We intuitively illustrate the concept of subspace expansion and rotation in Fig. \ref{Subspace_Demo}. Since the S\&C performance is determined by the inner product between the sensing (communication) subspace and the ISAC signal, the expanded subspaces obviously lead to better performance.
When the subspaces are nearly orthogonal, the minimum projection of the ISAC signal on the subspaces leads to poor reused power between S\&C.
By rotating the subspaces, the previously orthogonal S\&C subspaces are rotated and correlated to each other. 
Consequently, more signal power may be reused between S\&C leading to significantly improved ISAC performance.

To better interpret the gain provided by subspace rotation, we first define the correlation between the communication subspace and the sensing subspace as 
\begin{equation}
  \rho = \frac{\mathbf{h}_c^H\mathbf{h}_t}{\|\mathbf{h}_c\| \|\mathbf{h}_t\|}.
\end{equation}

Let us re-examine the optimal solution structure for a certain channel realization in Theorem \ref{Thm_optimal_beamformer}. 
If the absolute value of correlation is larger than $\sqrt{\Gamma_0 \sigma_c^2/(P_t \|\mathbf{h}_c\|^2)}$, the S\&C channels are strongly coupled, and most of the signal power can be reused by the dual functionalities. Thus, the beamformer aligned to the sensing subspace always satisfies the communication requirement. In this case, the communication performance can be written as
\begin{equation}
  \text{SNR}_c = \frac{P_t}{\sigma_c^2}\frac{|\mathbf{h}_c^H \mathbf{h}_t|^2 }{\|\mathbf{h}_t\|^2} = \frac{P_t}{\sigma_c^2} \|\mathbf{h}_c\|^2 |\rho|^2,
\end{equation}
which is increased with $\rho$.

If the absolute value of correlation is smaller than $\sqrt{\Gamma_0 \sigma_c^2/(P_t \|\mathbf{h}_c\|^2)}$, the S\&C channels are weakly coupled and communication SNR with the aforementioned beamformer may no longer fulfill the requirement.
In this case, the communication performance is always the same as the threshold, and the sensing performance can be written as
\begin{equation}
  \sqrt{\text{SNR}_s} =  \sqrt{\frac{\Gamma_0 \sigma_c^2}{\|\mathbf{h}_c\|^2}}|\rho|\|\mathbf{h}_t\| + \sqrt{P_t - \frac{\Gamma_0 \sigma_c^2}{\|\mathbf{h}_c\|^2}}\|\mathbf{h}_t\| \frac{(1-|\rho|)}{\sqrt{1-|\rho|^2}}.
\end{equation}
It is easy to verify that the sensing performance monotonously increases with  $\rho$ whenever $\rho$ is smaller than $\sqrt{\Gamma_0 \sigma_c^2/(P_t \|\mathbf{h}_c\|^2)}$.

\subsection{Proposed RIS-ISAC Beamforming Design}
As analyzed in the previous subsection, ISAC gain provided by RIS comes from subspace expansion and rotation.
It is natural to maximize the channel gain of each subsystem and the correlation between the S\&C channels by manipulating the phase shifters.
Given the RIS sensing and RIS communication channels $\mathbf{h}_r^*\mathbf{h}_t^H$ and $\mathbf{h}_c^H$ in (\ref{Tx_Channel})-(\ref{Comms_Channel}), the optimization problem for the subspace rotation and expansion algorithm can be formulated as
\begin{subequations}\label{heu}
  \begin{align}
    \min_{\bm{\Phi}} \quad &-\|\mathbf{h}_r^* \mathbf{h}_t^H \mathbf{h}_c\|^2\\
    s.t. \quad\ &|\varphi_m| = 1, \forall m = 1,2,...,M,\\
                &\bm{\Phi} = \operatorname{diag}(\varphi_1,....,\varphi_M).
  \end{align}
\end{subequations}
In (\ref{heu}), the objective is to maximize the inner product of the S\&C channels, which represents both channel correlation and channel gains.
To be mentioned, $\mathbf{w}$ is not included in this problem, since the subspace rotation and expansion method is not limited to any certain beamformer scheme.

This problem is non-convex due to both the non-convex objective function and the unit modulus constraints. For such a problem, we may apply the gradient projection algorithm to find a local optimum. 
For notational convenience, we denote $\mathbf{v} = \operatorname{diag}(\bm{\Phi})$, $\bar{\mathbf{B}} = \operatorname{DIAG}(\bar{\mathbf{b}})$, $\tilde{\mathbf{B}} = \operatorname{DIAG}(\tilde{\mathbf{b}})$ and $\mathbf{H}_{RU} = \operatorname{DIAG}(\mathbf{h}_{RU})$, where the $\operatorname{DIAG}$ operation transforms a vector to a diagonal matrix and $\operatorname{diag}$ operation transforms a diagonal matrix to a vector.
The objective function can be recast as
\begin{equation}
  f(\mathbf{v}) = -\Vert\mathbf{h}_r(\mathbf{v})\Vert^2 |\mathbf{h}_t^H(\mathbf{v})\mathbf{h}_c(\mathbf{v})|^2,
\end{equation}
where
\begin{align}
  &\mathbf{h}_t = \alpha_t (\mathbf{a}_t+\mathbf{G}_t \bar{\mathbf{B}}\mathbf{v}),\label{Tx_Channel_Impace} \\
  &\mathbf{h}_r = \alpha_r (\mathbf{a}_r+\mathbf{G}_r \tilde{\mathbf{B}}\mathbf{v}), \label{Rx_Channel_Impace}\\
  &\mathbf{h}_c = \mathbf{h}_{BU} + \mathbf{G}_t\mathbf{H}_{RU}\mathbf{v}\label{Comms_Channel_Impace}.
\end{align}
Thus, the gradient vector with respect to $\mathbf{v}$ can be written as 
\begin{equation}\label{Gradient}
   \nabla f(\mathbf{v}) = f_0 f_1 \nabla f_2(\mathbf{v}) + f_1 f_2 \nabla f_0(\mathbf{v}) + f_2 f_0 \nabla f_1(\mathbf{v}),
\end{equation}
where 
\begin{subequations}\label{Gradient_Each}
  \begin{align}
    &f_0 \!=\! -\Vert\mathbf{h}_r\Vert^2,       \nabla f_0\left(\mathbf{v})   \!=\! -|\alpha_r|^2(\mathbf{a}_r^H\mathbf{U}_t \!+\! \mathbf{v}^H\mathbf{U}_r^H\mathbf{U}_r\right),\\
    &f_1 = \mathbf{h}_t^H\mathbf{h}_c, \;\;  \nabla f_1(\mathbf{v})   =  \alpha_t^* \left(\mathbf{a}_t^H\mathbf{U}_c + \mathbf{v}^H\mathbf{U}_t^H\mathbf{U}_c\right),\\
    &f_2 = \mathbf{h}_c^H \mathbf{h}_t,  \;\;                   \nabla  f_2(\mathbf{v}) =    \alpha_t\left(\mathbf{h}_{BU}^H\mathbf{U}_t + \mathbf{v}^H\mathbf{U}_c^H\mathbf{U}_t\right),\\
    &\mathbf{U}_t= \mathbf{G}_{t}\bar{\mathbf{B}},   \quad
    \mathbf{U}_r = \mathbf{G}_{r}\tilde{\mathbf{B}}, \quad
    \mathbf{U}_c = \mathbf{G}_{t}\mathbf{H}_{RU}. \label{Define_U}
  \end{align}
\end{subequations}
The stepsize of each iteration is calculated using the backtracking line search method, and the projector is designed to normalize each element to have the unit modulus.

After optimizing the RIS, the optimal beamformer can be obtained by solving (\ref{Opt_prob_w}).
The overall algorithm is summarized as Algorithm \ref{alg_Heu}.

\renewcommand{\algorithmicrequire}{\textbf{Input}}
\renewcommand{\algorithmicensure}{\textbf{Output}}
\begin{algorithm}
\caption{Subspace rotation \& expansion scheme}
\label{alg_Heu}
\begin{algorithmic}
    \REQUIRE:
             $\alpha_r$, $\alpha_t$, $\theta$, $\phi$, $\mathbf{h}_{BU}$, $\mathbf{h}_{RU}$, $\mathbf{G}_r$, $\mathbf{G}_t$,$\Gamma_0$, $P_t$, $\bm{\Phi}$, $\sigma_c$, $\sigma_s$, maximal iteration $k_{max}$, tolerance $\varepsilon$.
    
    \ENSURE: The transmit beamformer $\mathbf{w}$.
    \STATE 1. Randomly initialize $\mathbf{v}$.
    \FOR {$k=1$ to $k_{max}$} 
    \STATE 2. Compute the gradient $\nabla f(\mathbf{v}^{(k-1)})$ in (\ref{Gradient}).
    \STATE 3. Update $\mathbf{v}$: $\mathbf{v}^{(k)} = \mathbf{v}^{(k-1)} - a \nabla f(\mathbf{v}^{(k-1)})$, where $a$ is the stepsize calculated by the backtracking line search method.
    \STATE 4. Project onto constraint set: $v_m = \frac{v_m}{|v_m|}$, $\forall m \in [1,M]$.
    \STATE \textbf{if}  $|f(\mathbf{v}^{(k)})-f(\mathbf{v}^{(k-1)})| < \varepsilon$, \textbf{break}.
    \ENDFOR
    \STATE 6. Calculate the optimal beamformer $\mathbf{w}$ in Theorem \ref{Thm_optimal_beamformer}.
\end{algorithmic}
\end{algorithm}

\section{Benchmark Scheme: Channel Gain Maximization}
To verify the effectiveness of the proposed subspace rotation method, we further develop a benchmark scheme that directly optimize the original optimization problem (\ref{Opt_Raw}).

Since the optimization variables $\bm{\Phi}$ and $\mathbf{w}$ are deeply coupled in (\ref{Opt_Raw}), the problem is non-convex and challenging to solve in general. 
To that end, we resort to alternatively optimizing the beamformer and the phase shift of each element to seek a local optimum.
Since the optimal beamformer has been fully investigated in Theorem 1, we focus on optimizing the phase shifters in this section.

\textbf{Closed-form Solution for the Phase Shifter}: To overcome the non-convexity of the phase shift matrix $\bm{\Phi}$, we decompose problem (\ref{Opt_Raw}) into multiple sub-problems, each of which optimizes an individual phase shifter while fixing the other ones.
The problem can be mathematically expressed as
\begin{subequations}\label{Optmize_Phi_AO}
  \begin{align}
    \max_{\varphi_m} \quad &\text{SNR}_s \label{obj_Phi} \\
    s.t. \quad\ &\text{SNR}_c \geq \Gamma_0,\label{Rate_constraint}\\
                &|\varphi_m| = 1.
  \end{align} 
\end{subequations}
Using (\ref{Tx_Channel_Impace})-(\ref{Comms_Channel_Impace}) and (\ref{Define_U}), the channel vectors can be decomposed into a more trackable form as
\begin{align}\label{impact_h}
  &\mathbf{h}_j = \alpha_j(\mathbf{a}_j + \mathbf{U}_j \mathbf{v}) \!=\! \alpha_j\Bigg(\Big(\mathbf{a}_{j} \!+\! \sum_{i=1,i\neq m}^M v_i\mathbf{u}_{j,i} \Big)\!+\! v_m \mathbf{u}_{j,m} \!\Bigg) \notag\\
  &= \alpha_j (\tilde{\mathbf{h}}_j + v_m \mathbf{u}_{j,m}), \forall j \in {t,r,c}
\end{align}
where $\alpha_c = 1$ and $\mathbf{u}_{j,m}$ denotes the $j$-th column of $\mathbf{U}_j,\; \forall j \in t,r,c$.

Based on the above decomposition, the objective function with respect to the $m$-th element of the phase shift matrix can be recast as
\begin{align}
  &\text{SNR}_s(v_m)  = \Vert  \mathbf{h}_r \Vert^2 |\mathbf{h}_t^H\mathbf{w}|^2\\
                &=\alpha_r^2 \alpha_t^2 \big(\Vert \tilde{\mathbf{h}}_r \Vert^2 +\Vert\mathbf{u}_{r,m}\Vert^2+ 2\text{Re}\{v_m \tilde{\mathbf{h}}_r^H\mathbf{u}_{r,m}\}\big) \notag \\
                &\times \big(|\tilde{\mathbf{h}}_t^H \mathbf{w}|^2 +|\mathbf{u}_{t,m}^H\mathbf{w}|^2+ 2\text{Re}\{v_m \tilde{\mathbf{h}}_t^H\mathbf{ww}^H\mathbf{u}_{t,m}\}\big)\\
                & = \alpha_r^2 \alpha_t^2 (K_{0} + 2\text{Re}\{v_m a_{0}\}) (K_{1}+2\text{Re}\{v_m a_{1}\})
\end{align}
where 
$K_{0} = \Vert \tilde{\mathbf{h}}_r \Vert^2 +\Vert\mathbf{u}_{r,m} \Vert^2$,
$K_{1} = |\tilde{\mathbf{h}}_t^H \mathbf{w}|^2 +|\mathbf{u}_{t,m}^H\mathbf{w}|^2$,
$a_{0} = \tilde{\mathbf{h}}_r^H\mathbf{u}_{r,m}$ and 
$a_{1} = \tilde{\mathbf{h}}_t^H\mathbf{ww}^H\mathbf{u}_{t,m}$.
Since $K_{0}$ represents the total strength of the direct link and the $M$-1 reflected links, while $a_{0}$ represents the strength of the $m$-th reflected link, $K_{0}$ is much larger than $a_{0}$ when $M\gg1$.
Similarly, $K_{1}$ is much larger than $a_{1}$ when $M \gg 1$ and the objective funciton can be approximated as $K_{0}K_{1} + 2K_{1}\text{Re}\{v_m a_{0}\} + 2K_{0}\text{Re}\{v_m a_{1}\}$.

\begin{thm}
  The angle of the optimal solution of the $m$-th element is one of the following four solutions
  \begin{subequations}\label{Thm_optmial_angle}
    \begin{align}
      &\mu_1 = \nu_{0} -\arctan(\frac{\sin(\nu_{1} \!-\! \nu_{0})|a_{1}|K_{0}}
      {|a_{0}|K_{1}+\cos(\nu_{1} \!-\! \nu _{0}) |a_{1}|K_{0}}),\\
      &\mu_2 = v_2+\pi,\\
      &\mu_{3,4} = \pm \operatorname{acos}(\frac{\Gamma_0 - |\tilde{\mathbf{h}}_c^H \mathbf{w}|^2 \!-\!|\mathbf{u}_{c,m}^H\mathbf{w}|^2 \!}{2|a_c|}) - \nu_c,
    \end{align}   
  \end{subequations}
   where $\nu_0 = \angle a_0$, $\nu_1 = \angle a_1$, $a_c = \tilde{\mathbf{h}}_c^H\mathbf{ww}^H\mathbf{u}_{c,m}$ and $\nu_c = \angle a_c$.
  \begin{equation}
    \mu^* = 
    \begin{cases}
      \mu_{1,2},\; \textup{if} \; \mu_{1,2}\; \textup{is feasible},\\
      \mu_{3,4},\; \textup{otherwise},
    \end{cases}
  \end{equation}
  where $\mu_{1,2}$ is the one which has the larger objective function value in $\mu_1$ and $\mu_2$, and $\mu_{3,4}$ is the one which has the larger objective function value in $\mu_3$ and $\mu_4$.

  Proof. \textup{See Appendix \ref{Proof_Thm_optmial_angle}}.
\end{thm}


By iteratively optimizing a phase shifter when the others are fixed, and optimizing the beamformer by using Theorem \ref{Thm_optimal_beamformer}, a local optimum of (\ref{Opt_Raw}) can be achieved.
For clarity, we summarize this channel gain maximization algorithm in Algorithm \ref{alg1}.
\renewcommand{\algorithmicrequire}{\textbf{Input}}
\renewcommand{\algorithmicensure}{\textbf{Output}}
\begin{algorithm}
\caption{Channel gain maximization benchmark scheme}
\label{alg1}
\begin{algorithmic}
    \REQUIRE:
             $\alpha_r$, $\alpha_t$, $\theta$, $\phi$, $\mathbf{h}_{BU}$, $\mathbf{h}_{RU}$, $\mathbf{G}_r$, $\mathbf{G}_t$, $\Gamma_0$, $P_t$, $\sigma_c$, $\sigma_s$, outer iteration tolerance $\varepsilon_k$, maximal outer iteration number $k_{max}$, inner iteration tolerance $\varepsilon_l$, maximal inner iteration number $l_{max}$.
    \ENSURE: The transmit beamformer $\mathbf{w}$.
    \STATE 1. Randomly initialize $\bm{\Phi}$.
    \FOR{$k=1$ to $k_{max}$}
    \STATE 2. Update the beamformer $\mathbf{w}$ by employing Theorem 1.
      \FOR{$l=1$ to $l_{max}$}
        \FOR{$m=1$ to $M$}
        \STATE 3. Update $\mu_m$ by employing theorem 2.
        \ENDFOR 
        \STATE \textbf{If} $|\text{SNR}_s^{(l)} - \text{SNR}_s^{(l-1)}| \leq \varepsilon_l$, \textbf{break}.
      \ENDFOR
        \STATE \textbf{If} $|\text{SNR}_s^{(k)} - \text{SNR}_s^{(k-1)}| \leq \varepsilon_k$, \textbf{break}.
    \ENDFOR
\end{algorithmic}
\end{algorithm}


\section{Numerical Results}

\begin{figure*}[!htbp]
  \subfigure [Tradeoff performance for different algorithms.]{
  \includegraphics [width=0.32\textwidth]{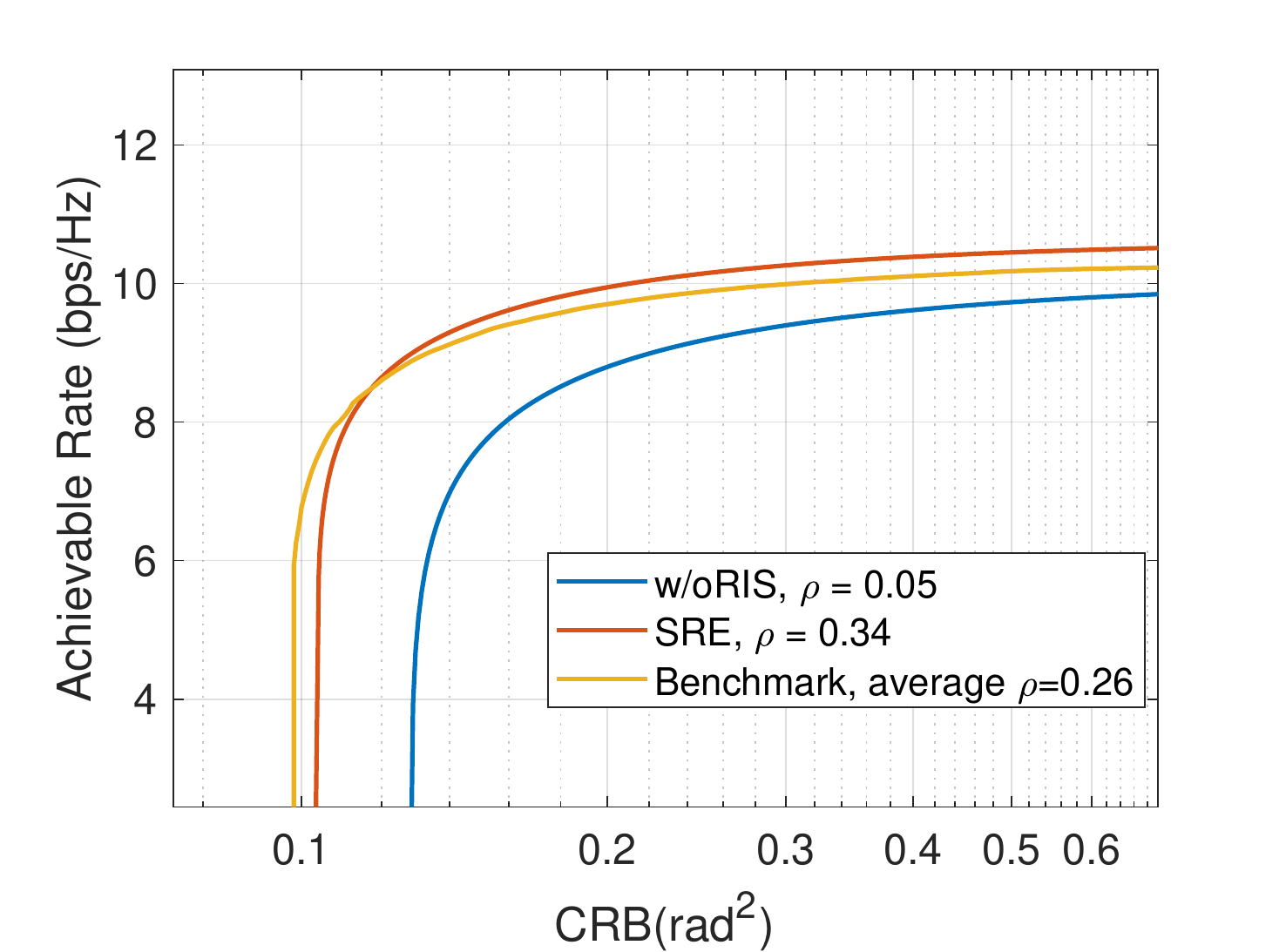}}
  \subfigure [CRB and $\rho$ v.s. size of RIS.]{
  \includegraphics [width=0.32\textwidth]{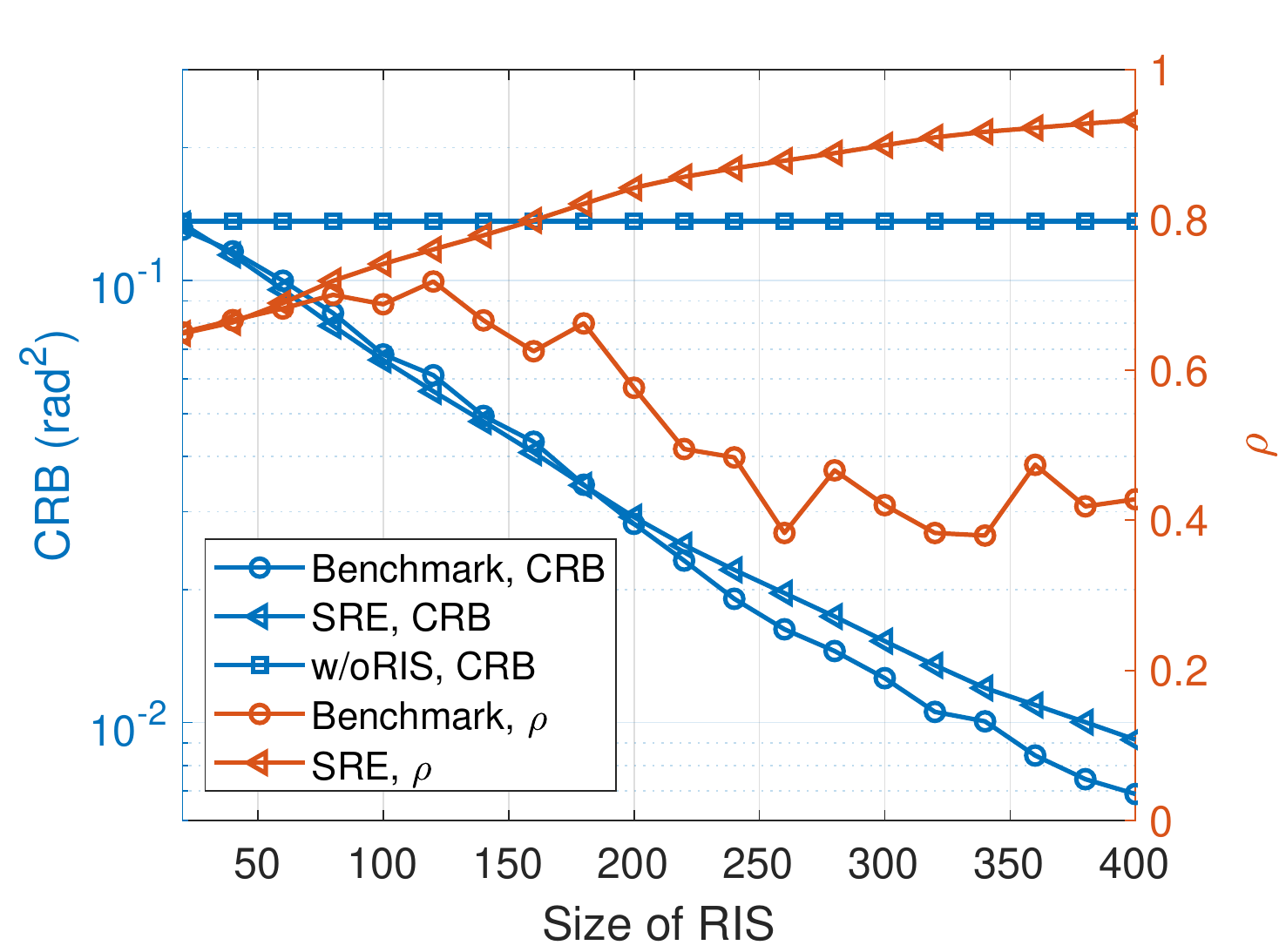}}
  \subfigure [Performace improvement derived from subspace rotation and expansion.]{
  \includegraphics [width=0.32\textwidth]{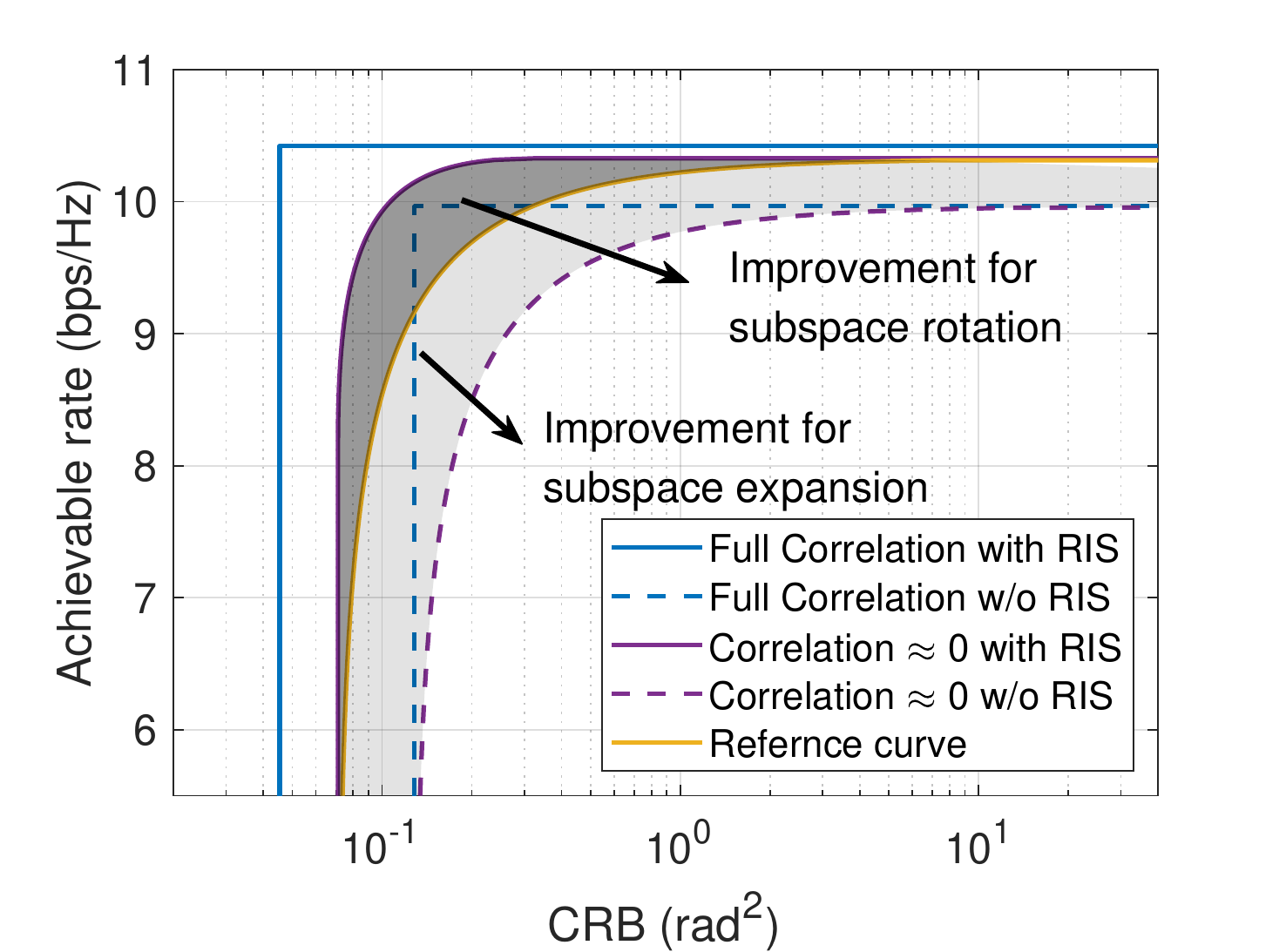}}
  \caption{Numerical results.}
  \label{NS}
\end{figure*}

In this section, we provide numerical simulations to validate the effectiveness of the proposed beamforming method and to illustrate the subspace rotation and expansion provided by RIS in the ISAC system.
The system parameters used in the simulations are summarized in the TABLE. \ref{Table}.
The channel gain maximization scheme is denoted as \textit{benchmark}, the subspace rotation and expansion scheme is denoted as \textit{SRE} and the ISAC system without the assistance of RIS is denoted as \textit{w/oRIS}.
\begin{table}[!htbp]
  \caption{Default System Parameter in the Simulations}
  \begin{tabular}{c c| c c} 
    \hline \hline
    Parameter         & Value   & Parameter            & Value\\
    \hline
    $P_t$             & 1W      &Location of the BS             &[0 0]m\\
    $\sigma_s^2$      & -60dBm  &Location of the RIS            &[30 30]m\\
    $\sigma_c^2$      & -60dBm  &Location of the Target         &[40 0]m\\
    $M$               & 64      &Distance from the BS to the UE & 30m\\
    $N_t$,$N_r$       & 15      &$f_c$             & 3GHz\\         
    \hline \label{Table}
  \end{tabular} 
\end{table}

In Fig. \ref{NS}(a), we show the tradeoff performance between the Cramer-Rao bound (CRB) of angle estimation with the communication rate of the three schemes, where the CRB is obtained from \cite{li2007range}.
It is obvious that the two RIS-assisted schemes have better tradeoff performance than that of w/oRIS.
Since the objective functions are different, there is a cross-over of the tradeoff curve between the benchmark algorithm and the SRE algorithm. 
However, the comparable performance verifies the effectiveness of rotating and expanding the S\&C subspaces, which has a significantly lower complexity.
Moreover, the improved channel correlation of the benchmark scheme shows the importance of channel correlation even if it is neither included in the objective function nor the constraints.

In Fig. \ref{NS}(b), we investigate the sensing performance with respect to the increased size of RIS under the given communication threshold. With more elements, the RIS has a stronger ability to modify the signal propagation environment, which leads to better sensing performance with given communication requirements. It is worth noting that although the two schemes have similar performance, the ways to improve the ISAC gain are different. The benchmark scheme tries to enhance the channel gain of each subsystem while the SRE algorithm invests more efforts in improving the channel correlation. 

In Fig. \ref{NS}(c), we differentiate the impacts brought by the subspace rotation and expansion to provide more insights into the proposed algorithm.
The dashed line shows the fully correlated ISAC system without the assistance of RIS can simultaneously achieve the best communication and sensing performance, while the tradeoff between them in the non-correlated system is obvious. This indicates that channel correlation is the decisive factor of integration gain when the strength of each channel is constant. Besides, the area between the two purple lines can be regarded as the performance gain of employing the RIS. 
To distinguish between the gains of subspace rotation and expansion, the reference curve with enhanced channel gain but zero S\&C channel correlation is provided. 
Compared with the RIS-assisted system with low correlation, the reference curve has the same minimal CRB and maximal achievable rate but different tradeoff performance.
Since the S\&C channel of the reference curve is still non-correlated, the light shadow below the reference curve represents the performance improvement for the enhanced channel gain, namely subspace expansion. 
Similarly, since the channel gain is the same as the solid purple line, the performance gain in the dark shadow comes from the subspace rotation.

\begin{table}[!htbp]
  \centering
  \caption{Computational overhead for different schemes.}
  \begin{tabular}{c c| cc c} 
    \hline 
    Scheme              &   & w/oRIS      & SRE            &benchmark\\
    \hline
    Execution time (s)   &   & 0.946       &4.73            &36.9 \\\hline
    \label{Table2}
  \end{tabular} 
\end{table}

Finally, we show the numerical computational complexity for 10000 channel realizations in TABLE \ref{Table2}.
It can be observed in TABLE \ref{Table2} that, due to the double loop, the benchmark scheme has a much larger computational overhead than that of the SRE method.
Moreover, since the SRE method is decoupled with transmit beamforming design problem in (\ref{Opt_prob_w}), the flexible tradeoff between S\&C can be achieved by leveraging the optimal ISAC beamformer.
The benchmark scheme, on the other hand, needs to be executed iteratively when the communication requirement changes.
\section{conclusion}
In this paper, we proposed two joint active and passive beamforming designs for RIS assisted ISAC system which serves a single antenna user and tracks a single target. 
In particular, we first pointed out that the integration gain of an ISAC system originates from the correlation between the communication and sensing subspaces by analyzing the structure of the optimal beamformer.
In light of this, we proposed a low-complexity heuristic algorithm to rotate and expand the S\&C subspaces.
We further formulated an optimization problem that maximizes the sensing SNR while guaranteeing communication SNR to provide a performance baseline.
Since the considered problem is non-convex, we derived the closed-form optimal solution for the subproblems.
Numerical results are provided to validate the effectiveness of the proposed schemes, which confirmed the performance gain brought by of subspace rotation and expansion.

\appendices

\section{Proof of Theorem 2}\label{Proof_Thm_optmial_angle}
Since the objective function is periodic about $\angle v_m$, we focus on the domain where $0 \leq \angle v_m \leq 2\pi$.
The objective function can be rewritten as
\begin{equation}
  g(\mu) = K_0 K_1 + 2K_1 \cos(\mu + \nu_0) + 2K_0 cos(\mu+\nu_1),
\end{equation}
where $\mu = \angle v_m$.
The derivative of the objective function is 
\begin{equation}
  \frac{\textup{d}g}{\textup{d}\mu} = -2K_1 \sin(\mu+\nu_0) - 2K_1 \sin(\mu + \nu_1).
\end{equation}
By checking the derivative, the objective function can be proved to have two extreme points, which are
\begin{subequations}\label{opt_phase_shift}
  \begin{align}
    &\mu_1 =\nu_{0} -\arctan\left(\frac{\sin(\nu_{1} \!-\! \nu_{0})|a_{1}|K_{0}}
                              {|a_{0}|K_{1}+\cos(\nu_{1} \!-\! \nu _{0}) |a_{1}|K_{0}}\right)\\
    &\mu_2 =\pi \!+\! \nu _{0} -\arctan\left(\frac{\sin(\nu _{1} \!-\! \nu_{0})|a_{1}|K_{0}}
    {|a_{0}|K_{1}+\cos(\nu_{1} \!-\! \nu _{0}) |a_{1}|K_{0}}\right). \label{opt_obj}
  \end{align}
\end{subequations}
Since the objective is periodic, the two extream points must respectively be the largest one and the smallest one.

Then we discuss the feasible problem in (\ref{Rate_constraint}).
The SNR at the user can be recast as
\begin{align}
\text{SNR}_c& = |\mathbf{h}_c^H \mathbf{w}|^2 \notag \\
               &= |\tilde{\mathbf{h}}_c^H \mathbf{w}|^2 \!+\!|\mathbf{u}_{c,m}^H\mathbf{w}|^2 \!+\! 2\text{Re}\{v_m \tilde{\mathbf{h}}_c^H\mathbf{ww}^H\mathbf{u}_{c,m}\}.
\end{align}
To meet the communication requirement, $\mu$ should be satisfied the following condition
\begin{equation}
  \cos(\mu+\nu_c) \geq \frac{\Gamma_0\sigma_c^2 - |\tilde{\mathbf{h}}_c^H \mathbf{w}|^2 \!+\!|\mathbf{u}_{c,m}^H\mathbf{w}|^2}{2|a_c|}.
\end{equation}
Thus the two bounds of the feasible region are
\begin{align}
  &\mu_{3} = \operatorname{acos}(\frac{\Gamma_0\sigma_c^2 - |\tilde{\mathbf{h}}_c^H \mathbf{w}|^2 \!-\!|\mathbf{u}_{c,m}^H\mathbf{w}|^2 \!}{2|a_c|}) - \nu_c,\\
  &\mu_{4} = -\operatorname{acos}(\frac{\Gamma_0\sigma_c^2 - |\tilde{\mathbf{h}}_c^H \mathbf{w}|^2 \!-\!|\mathbf{u}_{c,m}^H\mathbf{w}|^2 \!}{2|a_c|}) - \nu_c,
\end{align}
and the feasible region can be epxressed as
\begin{equation}
  \mu \in [\mu_4 - \mu_c , \mu_3 - \mu_c ].
\end{equation}
If the larger one of $\mu_1$ and $\mu_2$ is feasible, the optimal solution is obviously the $\mu_1$ or $\mu_2$, otherwise, it is easy to prove that the optimal solution should on the bound of the feasible region, i.e., $\mu_3$ or $\mu_4$.


\ifCLASSOPTIONcaptionsoff
  \newpage
\fi



\bibliographystyle{IEEEtran}
\bibliography{IEEEabrv,CEP_REF,EKF_RG}
\end{document}